\documentclass[aps,prb,superscriptaddress,twocolumn,10pt]{revtex4-2}

\usepackage{graphicx}
\usepackage{dcolumn}
\usepackage{bm}
\usepackage{amsfonts}
\usepackage{tikz}

\begin{document}

\preprint{APS/123-QED}

\title{Two-Photon Interference from Silicon-Vacancy Centers in Remote Nanodiamonds}
\author{R. Waltrich}
\affiliation{Institute for Quantum Optics, Ulm University, 89081 Ulm, Germany}
\author{M. Klotz}
\affiliation{Institute for Quantum Optics, Ulm University, 89081 Ulm, Germany}
\author{V. N. Agafonov}
\affiliation{GREMAN, UMR 7347 CNRS, INSA-CVL, Tours University, 37200 TOURS, France}
\author{A. Kubanek}
\affiliation{Institute for Quantum Optics, Ulm University, 89081 Ulm, Germany}

\date{\today}

\begin{abstract}
The generation of indistinguishable photons is a key requirement for solid-state quantum emitters as a viable source for applications in quantum technologies. Restricting the dimensions of the solid-state host to a size well below the wavelength of light emitted by a defect-center enables efficient external optical coupling, for example for hybrid integration into photonic devices. However, stringent restrictions on the host dimensions result in severe limitations on the spectral properties reducing the indistinguishability of emitted photons.  Here, we demonstrate two-photon interference from two negatively-charged Silicon-Vacancy centers located in remote nanodiamonds. The Hong-Ou-Mandel interference efficiency reaches 61\% with a coalescence time window of 0.35 ns. We furthermore show a high yield of pairs of Silicon-Vacancy centers with indistinguishable optical transitions. Therefore, our work opens new paths in hybrid quantum technology based on indistinguishable single-photon emitters in nanodiamonds. 
\end{abstract}

\maketitle
In the emerging field of quantum technologies the distribution of entanglement is a key ingredient, for example to establish long-distance quantum state transfer and quantum networks \cite{Wehner2018}. One possible source of distributed entanglement generation is two-photon interference (TPI), commonly known as Hong-Ou-Mandel (HOM) interference \cite{HongOuMandel1987}. A prerequisite are single photon sources that produce indistinguishable photons. Two-photon interference has been demonstrated with various sources of single photons, for example atomic vapors \cite{Chaneliere2007}, quantum dots \cite{Flagg2010}, molecules \citep{Lettow2010}, coupled atom-cavity systems \cite{Legero2004} and negatively-charged Nitrogen-Vacancy (NV$^-$) centers in bulk-diamond \citep{Bernien2012, Sipahigil2012}. Group IV color centers in diamond and, in particular, the negatively-charged Silicon-Vacancy center (SiV$^-$) are of great interest for the generation of indistinguishable photons, due to intrinsic spectral stability and narrow inhomogeneous line-distribution shown for bulk diamond \citep{Rogers2014, Jahnke2014}. In recent years, there is an increasing effort to restrict the dimensions of the diamond host to very small size well below the optical wavelength. Such nanodiamonds (NDs) give rise to a large variety of new applications in the realm of hybrid quantum systems. Individually optimized \citep{Fehler2021, Bayer2022} and large-scale \citep{Wan2020, Schrinner2020} hybrid quantum photonic circuits \citep{Elshaari2020, Kubanek2022, Sahoo2022} are constructed by means of nanomanipulation \citep{Hausler2019} and advanced fabrication techniques, respectively. Initially, the spectral properties of color centers in NDs were inferior compared to bulk diamond blocking further use in quantum optics applications. This deficiency was resolved over the past years by improved ND-production, sample preparation and control techniques \citep{RogersWang_2019}. \\
In this letter, we demonstrate two-photon interference from SiV$^-$ in remote NDs. Together with recently demonstrated access to electron spin \citep{Klotz2022} this work marks a further step towards applied hybrid quantum technology, such as the realization of scalable quantum networks, based on SiV$^-$ in NDs.

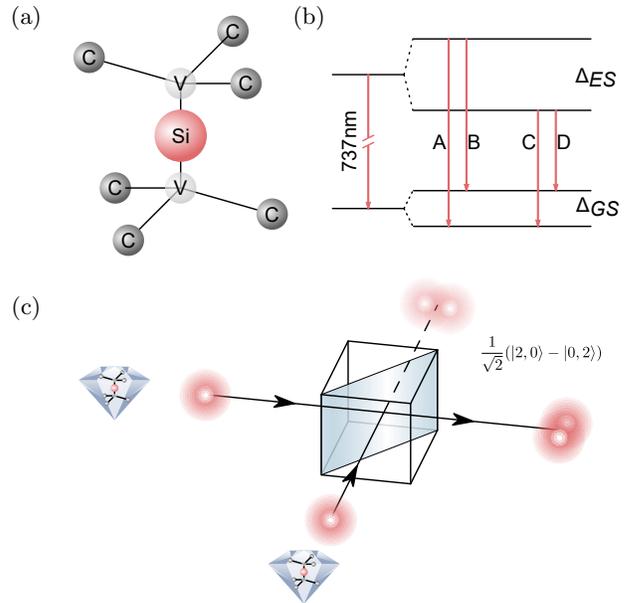
\begin{figure}[t]
	\begin{tikzpicture}
	\draw (0,0) node[inner sep=0] {\includegraphics[scale=0.95]{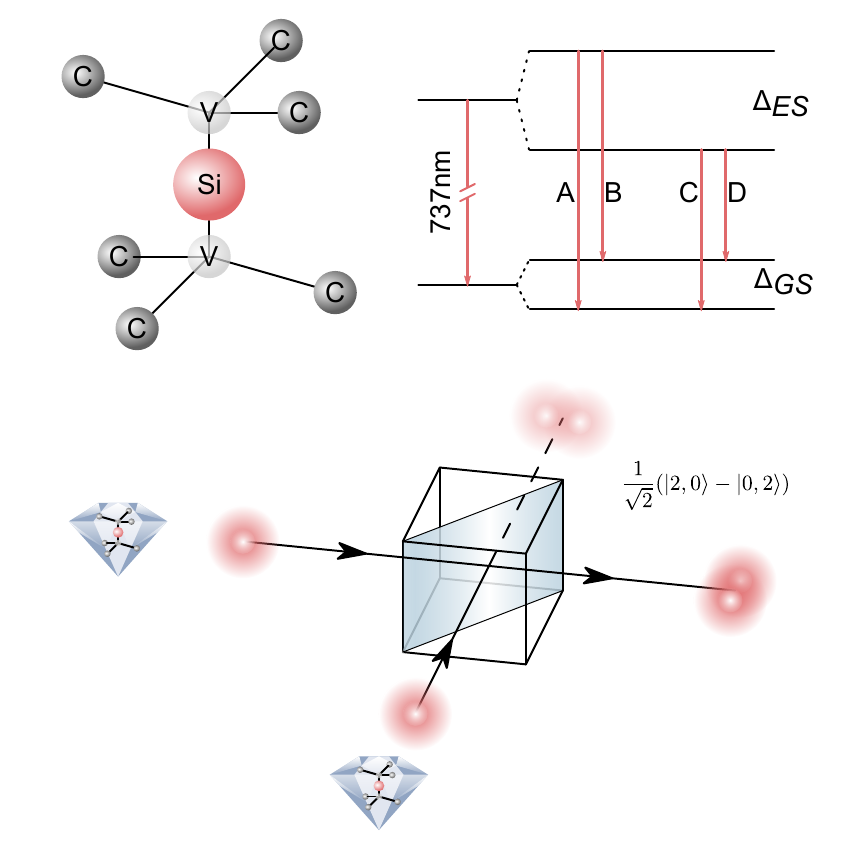}};
	\draw (-117pt, 110pt) node {(a)};
	\draw (-10pt, 110pt) node {(b)};
	\draw (-117pt, 0pt) node {(c)};
	\end{tikzpicture}
\caption{\label{fig:Fig1} a) Molecular structure of the SiV$^-$. A silicon atom (Si) is located between two adjacent carbon vacancies (V) along the [111] axis of the diamonds crystal structure. b) Electronic level structure of the SiV$^-$ with four optical transitions A, B, C, D arising from the ground and excited state doublets ($\Delta_{GS}$, $\Delta_{ES}$) due to spin-orbit interaction. c) Schematic of the two-photon interference experiment. Two identical photons from two separate SiV$^-$ enter a beam splitter from two different input ports. Constructive interference leads to maximal coalescence at the output ports.}
\end{figure}

The SiV$^-$ is a point defect in the lattice of a diamond crystal where a silicon atom is located between two adjacent carbon vacancies as depicted in FIG. \ref{fig:Fig1} a). At cryogenic temperatures four optically active transitions, resulting from spin-orbit coupling, can be observed. We refer to them as transitions A, B, C, D, as depicted in FIG. \ref{fig:Fig1} b). 
To show two-photon interference, we excited two SiV$^-$ in two remote NDs, separated by approximately 95 $\mu$m, off-resonantly and spectrally filtered the dominant transition C. The photons from both SiV$^-$ interfered on a 50:50 beamsplitter, as schematically shown in FIG. \ref{fig:Fig1} c). In the case of two identical photons entering the beamsplitter from two different input ports, the probability amplitudes for leaving at the same port will interfere constructively while the ones for leaving at different output ports interfere destructively. A second-order correlation measurement will therefore result in antibunching with vanishing correlations at zero time delay $g^{(2)}(\tau = 0) = 0 $. In contrast, $g^{(2)}(\tau = 0)=0.5$ is indicative of interference of two single but distinguishable photons.

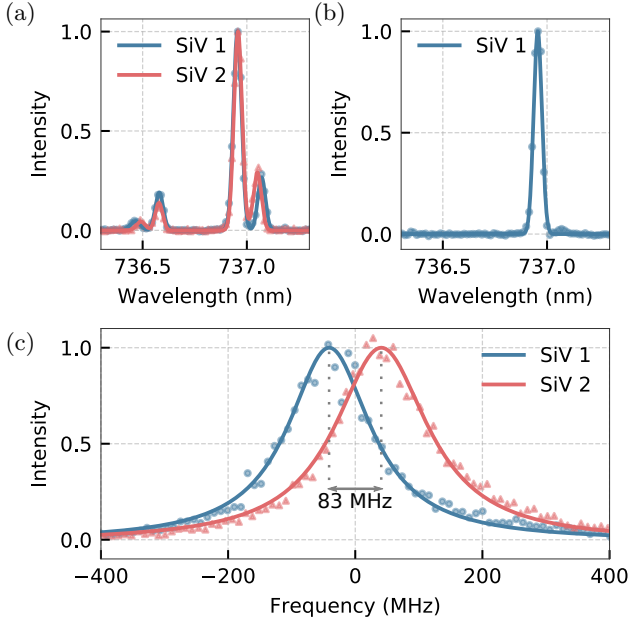
\begin{figure}[t]
	\begin{tikzpicture}
	\draw (0,0) node[inner sep=0] {\includegraphics[scale=0.95]{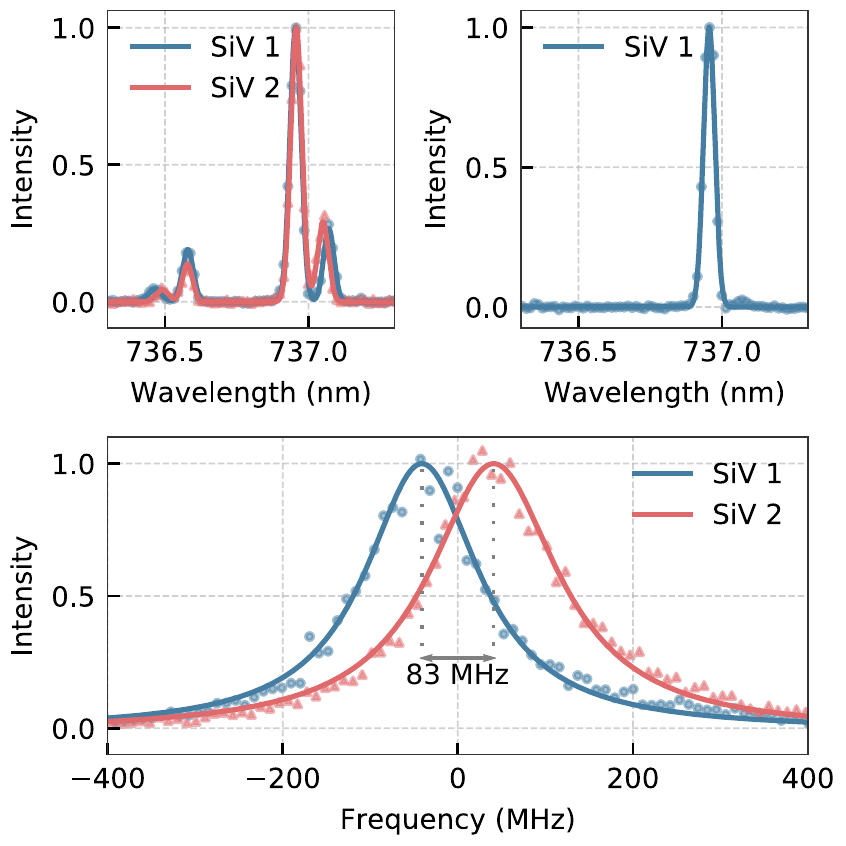}};
	\draw (-117pt, 115pt) node {(a)};
	\draw (0pt, 115pt) node {(b)};
	\draw (-117pt, -10pt) node {(c)};
	\end{tikzpicture}
\caption{\label{fig:Fig2} a) PL spectra of SiV 1 and SiV 2 (blue and red). The measured wavelength of transition C is identical for both color centers. They differ only in their ground state splitting (64 GHz and 52 GHz) and local temperature (7.7K and 6.2K). b) PL spectrum of SiV 1 after filtering with the etalons, leaving only transition C visible. c) Normalized PLE scans of SiV 1 and SiV 2 (red, blue) with measured respective linewidths of ($158 \pm 5$) MHz and ($177 \pm 4$) MHz and a detuning of ($83 \pm 6$) MHz. The plot is centered around their common center at 406.827829 THz.}
\end{figure}

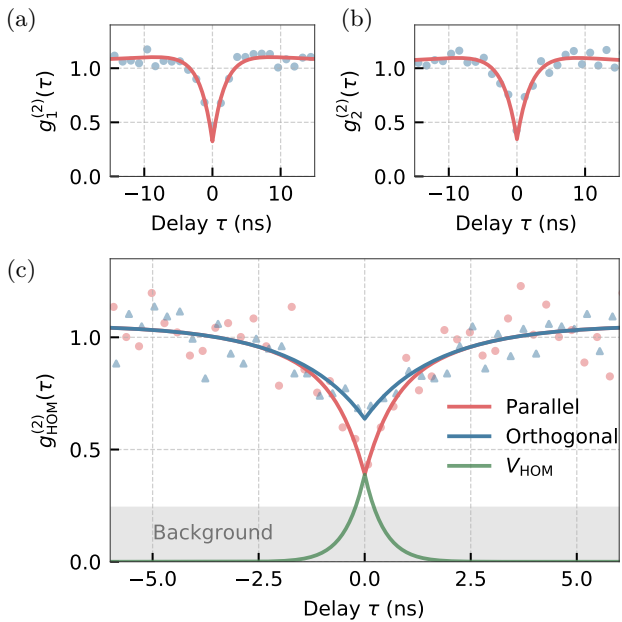
\begin{figure}[t]
	\begin{tikzpicture}
	\draw (0,0) node[inner sep=0] {\includegraphics[scale=0.95]{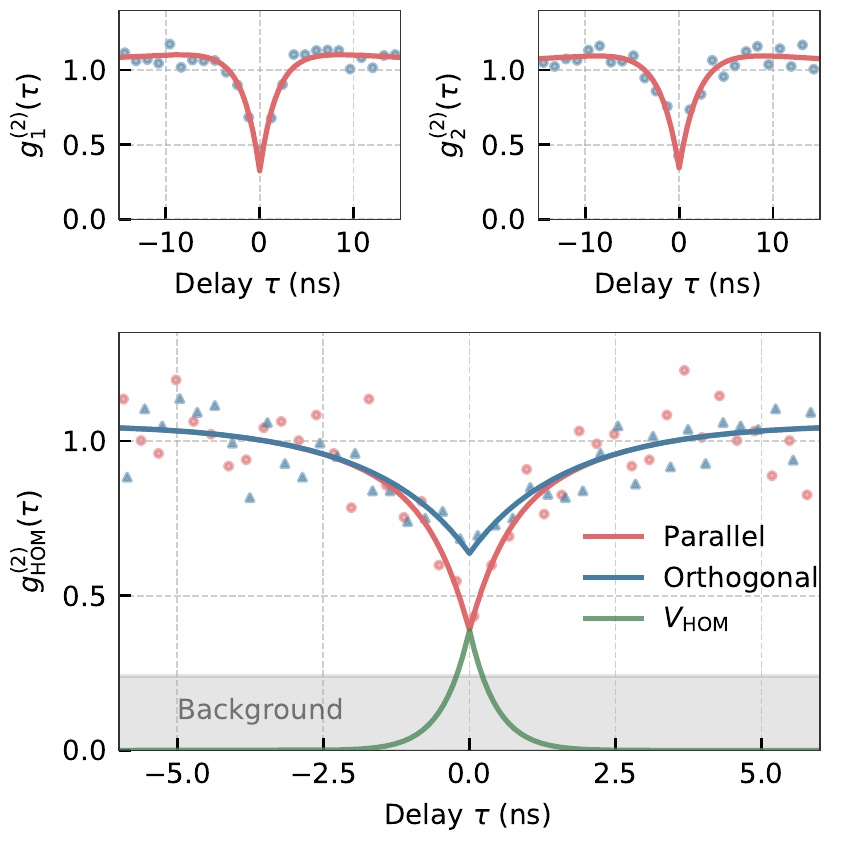}};
	\draw (-117pt, 115pt) node {(a)};
	\draw (10pt, 115pt) node {(b)};
	\draw (-117pt, 20pt) node {(c)};
	\end{tikzpicture}
\caption{\label{fig:Fig3} a) Normalized correlation function of SiV 1 with $g_1^{(2)}(0) = 0.33 \pm 0.07 $ and $\tau_1 = (1.84 \pm 0.29)$ ns. b) Normalized correlation function of SiV 2 with $g_2^{(2)}(0) = 0.35 \pm 0.08 $ and $\tau_2 = (2.02 \pm 0.35)$ ns. c) Two-photon interference between SiV 1 and SiV 2 with $\eta=0.61\pm 0.16$. Parallel polarization, corresponding to indistinguishable photons, is shown with red data points. Orthogonal polarization, corresponding to distinguishable photons, is depicted by blue triangles. The visibility $V_\mathrm{HOM}$ of the interference is shown in green.}
\end{figure}

The SiV$^-$s used for the experiment are located inside NDs with an avarage diameter of around 30 nm. They were coated onto a diamond substrate to ensure good thermal conductivity. The sample was investigated by off-resonant photo-luminescence (PL) and resonant photo-luminescence-excitation (PLE) measurements showing predominantly single SiV$^-$ and a spectral distribution of $\approx$ 14 GHz for transition C (see Appendix B). To find a matching SiV$^-$ pair we fixed the frequency of the scanning laser to the resonance of transition C of one SiV$^-$ and scanned the sample laterally. This way only SiV$^-$s with spectral overlap were visible. 
The spectra of the two SiV$^-$ chosen for the HOM measurement are shown in FIG. \ref{fig:Fig2} a). Their ground-state splitting (GSS) differs by 12 GHz, but transition C shows a good overlap. The fact that transition C overlaps although the GSS of both emitters differs can be explained by different combinations of axial and transverse strain of the host crystal \citep{Meesala2018}. After filtering transition C (see Appendix A for details), only a single line was observed for each SiV$^-$, as exemplary shown for SiV 1 in FIG. \ref{fig:Fig2} b). PLE measurements of the C transitions for both SiV$^-$s revealed linewidths of ($158 \pm 5$)  MHz and ($177 \pm 4$)  MHz with a detuning $\frac{\Delta}{2 \pi}$ of ($83 \pm 6$) MHz, as shown in FIG. \ref{fig:Fig2} c). Single-photon emission from both SiV$^-$s was confirmed by off-resonant second-order correlation measurements which, after a normalization, resulted in $g^{(2)}_1(0) = 0.33 \pm 0.07$ and $g^{(2)}_2(0) = 0.35 \pm 0.08$ as depicted in FIG. \ref{fig:Fig3} a) and b). We use the notation $g^{(2)}_i(\tau)$ for correlation functions including background noise, while the calligraphic $\mathfrak{g}^{(2)}(\tau)$ is used for the modeled correlation function without background. The measured data was fitted with 

\begin{equation}
	g_i^{(2)}(t) = 1 + \frac{S_i^2}{I_i^2}\mathfrak{g}_i^{(2)}(t) \ ,
\end{equation}

where $S_i$ is the signal from the emitter $i$, $I_i = S_i + B_i$ is the total signal including background counts $B_i$ and $\mathfrak{g}_i^{(2)}(t) = 1 - (1+a) \cdot \mathrm{exp}(-|t| / \tau_1) + a \cdot  \mathrm{exp}(-|t| / \tau_2)$ is a three level model of the correlation function \citep{Aharonovich2010}. From the fit we determined the signal to noise ratio $S_i / B_i \approx 4$ for each individual SiV$^-$ which sets a lower bound for the expected HOM dip, illustrated by the gray area in FIG. \ref{fig:Fig3} c).
To measure the two-photon interference we then off-resonantly excited both SiV$^-$s independently and let the emitted photons interfere on a 50:50 beamsplitter after the polarization of the photons was matched by half-wave plates. The resulting correlation function is shown in FIG. \ref{fig:Fig3} c), where red dots show data for parallel polarization and blue triangles show data for perpendicular polarization. The data was fitted with \citep{Lettow2010}

\begin{eqnarray}
g_{\mathrm{HOM}}^{(2)}(\tau) = && \ c^2_1 g^{(2)}_{1}(0) + c^2_2 g^{(2)}_{2}(0) + 2c_1 c_2 \nonumber \\
&& \ \cdot \left( 1 - \eta \frac{\langle S_1 \rangle \langle S_2 \rangle}{\langle I_1 \rangle \langle I_2 \rangle} |g^{(1)}_1(\tau)||g^{(1)}_2(\tau)| \cos(\Delta \tau) \right), \nonumber \\
\end{eqnarray}

where $c_i = I_i / (I_1 + I_2)$ and $g^{(1)}_i(\tau) = \exp(-\gamma_i|\tau|/2)$. 
The signal and noise for each emitter were fixed with the previously determined signal to noise ratio of the individual correlation measurements of both SiV$^-$. The variable $\eta$ in front of the interference term can be interpreted as an efficiency coefficient, where a value of 0 means no two-photon-interference and 1 means maximum interference. We determined a value of $\eta=0.61\pm 0.16$ for the case of parallel polarization. 

As an additional figure-of-merit we calculated the coalescence time window (CTW) as described in \citep{Proux2015}. It gives a time-window for which coalescence can occur on the beam-splitter. By integrating over the visibility function
\begin{eqnarray}
	V_{\mathrm{HOM}} = 1 - g_{\parallel}^{(2)}(\tau) / g_{\perp}^{(2)}(\tau)
\end{eqnarray}
we find
\begin{eqnarray}
\mathrm{CTW} = \int V_{\mathrm{HOM}} \mathrm{d}\tau = 0.35 \ \mathrm{ns}
\end{eqnarray}
which is below the limit of 2$T_1$ expected for two ideal emitters with absolutely indistinguishable photons. Here $T_1 \approx 1.9$ ns is the excited state lifetime. The result is limited by dephasing of the the emitters and background noise. \\
We demonstrated the generation of indistinguishable single photons from SiV$^-$ in two remote NDs with an efficiency of $\eta=0.61$, the extracted CTW yielded 0.35 ns. The interference visibility is limited by technical imperfections such as polarization drifts or detector timing response, which can be improved in future experiments. Also minor spectral diffusion during the measurement can diminish the visibility. Our results establish SiV$^-$ in NDs as a viable source for the generation of indistinguishable photons and open new possibilities for the integration into hybrid quantum photonics. The incorporation into photonic structures boosts the operation bandwidth and paves the way to establish remote entanglement of distant quantum nodes in an integrated fashion. \\
Furthermore, NDs much smaller than the wavelength of light, which contain individual quantum emitters with spectrally indistinguishable transitions offer new possibilities for the construction of cooperative quantum materials. Spatial indistinguishability can be achieved by positioning the NDs in a collective mode within a volume small compared to the third power of the radiation wavelength, $V \ll \lambda ^3$. Thereby, collective states can be prepared in the Dicke regime \citep{Dicke1954} in a bottom-up approach by means of AFM-based nanomanipulation. Cooperative processes such as superradiance \citep{GROSS1982} and superabsorption \citep{Higgins2014} can be accessed with color centers in diamond. Pioneering work demonstrated superradiance effects with ensembles of NV$^-$ center in the optical \citep{Bradac2017, Gutsche2022} and microwave \citep{Angerer2018} domain and indicated a first onset of cavity-assisted superabsorption \citep{Hausler2019b}. With the presented work, collective states could now be engineered atom-by-atom and with altered settings from spatially confined, interacting atoms to far distant non-interacting atoms \citep{Bojer2022}. 

\begin{acknowledgments}
We thank V. A. Davydov for the synthesis of the nanodiamonds. The project was funded by the Baden-W\"urttemberg Stiftung in the project Internationale Spitzenforschung. A.K. acknowledges support of the BMBF/VDI in the project QR.X and Spinning.

\end{acknowledgments}
\appendix
\section{Methods}
The nanodiamonds used for the experiment where synthesized by high pressure high temperature treatment of a mixture of Naphthalene - C$_{10}$H$_8$ and the Si doping component Tetrakis(Trimethylsilyl)silane – C$_{12}$H$_{36}$Si$_5$. The average size is around 30nm and the NDs mostly host single SiV$^-$. The same NDs where used in \citep{Klotz2022}.
\begin{figure}[ht]
\includegraphics{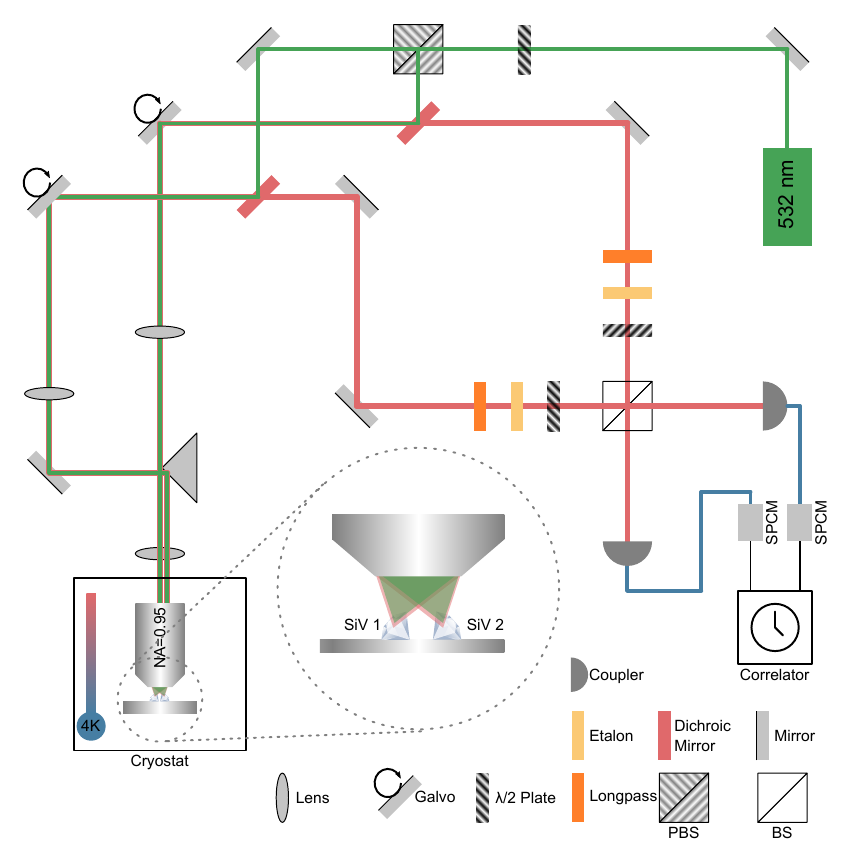}
\caption{\label{fig:Setup}Schematic of the optical setup for the two-photon interference experiment. A continious-wave laser with 532nm was used for excitation of the individual color centers SiV 1 and SiV 2 inside a flow-cryostat. The fluorescence was directed through spectral filters, two consecutive etalons and half-wave plates and interfered at a 50:50 non-polarizing beam splitter. The photons were collected through single-mode fibers into SPCMs.}
\end{figure}

To measure TPI we used a home-built, two-channel confocal microscope as depicted in FIG. \ref{fig:Setup}. A continuous-wave 532nm laser was used to excite the individual color centers off-resonantly. The beam was split at a 50:50 polarizing beamsplitter and a half-wave plate before the beam splitter allowed to adjust the excitation power for each channel and thus balance the emission from both SiV$^-$. Two galvo-scanners were used to direct the beam onto the SiV$^-$. A knife edge prism divided the field of view of the confocal setup into two independent channels. The sample was placed inside a continuous-flow cryostat and cooled with liquid helium to around 4K. The NDs were coated onto a diamond substrate for good thermal conductivity and reached local temperatures between 5K and 10K. Fluorescence of the SiV$^-$ in both channels was filtered with a dichroic mirror, a 740/13 band-pass filter and two consecutive etalons. The first etalon has a free spectaral range (FSR) of 850 GHz and a linewidth of 90 GHz, the second one has a FSR of 10 GHz and a linewidth of 1 GHz. Before both paths were joined on a 50:50 non-polarizing beamsplitter, a half-wave plate was used to adjust their polarization. Photons were then collected with two single mode fibers, detected by single photon counting modules (SPCM) and correlated with a time tagger device.

\begin{figure}[ht]
	\begin{tikzpicture}
	\draw (0,0) node[inner sep=0] {\includegraphics[scale=0.95]{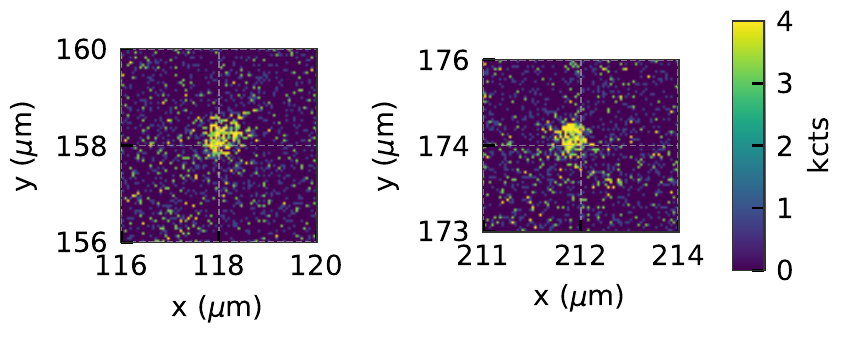}};
	\draw (-117pt, 40pt) node {(a)};
	\draw (-17pt, 40pt) node {(b)};
	\end{tikzpicture}
\caption{\label{fig:confocal}Confocal scan of a) SiV 1, and b) SiV 2. Both SiV$^-$ are spatially isolated from other emitters and separated by 95 $\mu$m.}
\end{figure}
A confocal-scan of both SiV$^-$ used for the TPI is shown in FIG.  \ref{fig:confocal}. Each SiV$^-$ was spatially isolated from other SiV$^-$ and both are spatially separated by 95 $\mu$m. 

\begin{figure}[t]
	\begin{tikzpicture}
	\draw (0,0) node[inner sep=0] {\includegraphics[scale=0.95]{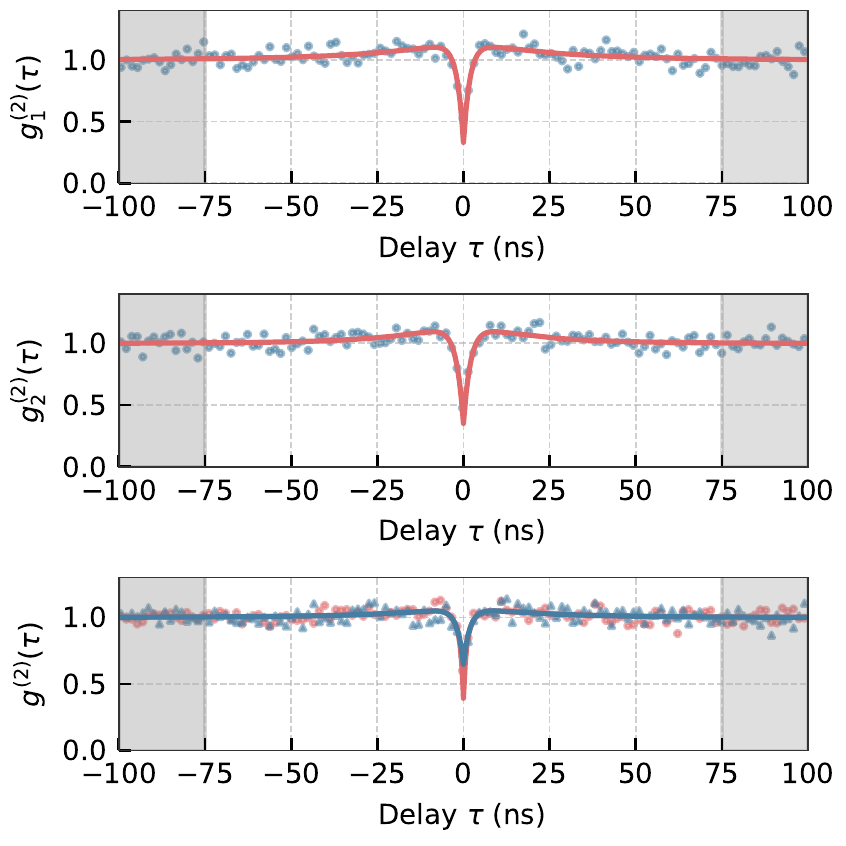}};
	\draw (-120pt, 115pt) node {(a)};
	\draw (-120pt, 33pt) node {(b)};
	\draw (-120pt, -50pt) node {(c)};
	\end{tikzpicture}
\caption{\label{fig:g2_full_range}Long time dynamics of the measured correlation functions of SiV 1 and SiV 2 (a), b)) and the two-photon-interference (c)), revealing a minor bunching behavior. The coincidence levels within the gray shaded regions where used to normalize the correlation functions.}
\end{figure}
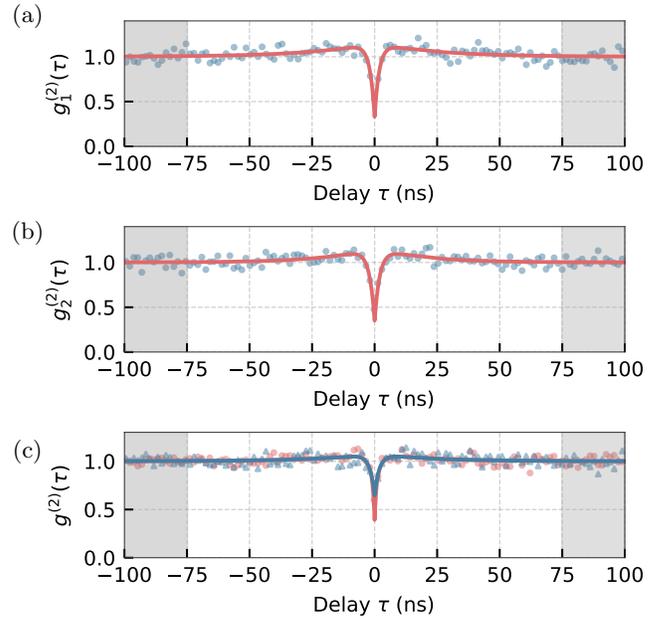

To determine the long time dynamics of the emitters and ensure a correct normalization, we analyzed the correlation functions of both emitters and fitted them to a three level model for the correlation function. Both SiV$^-$ showed a small amount of bunching and a shelving time of around 25 ns. Therefore, the correlation functions $g^2(\tau)$ are normalized to 1 with the averaged value of $g^2(\tau)$ within $75$ ns $<$ $\tau < 100$ ns as highlighted by a grey areas in FIG. \ref{fig:g2_full_range}.

\section{Properties of the SiVs}
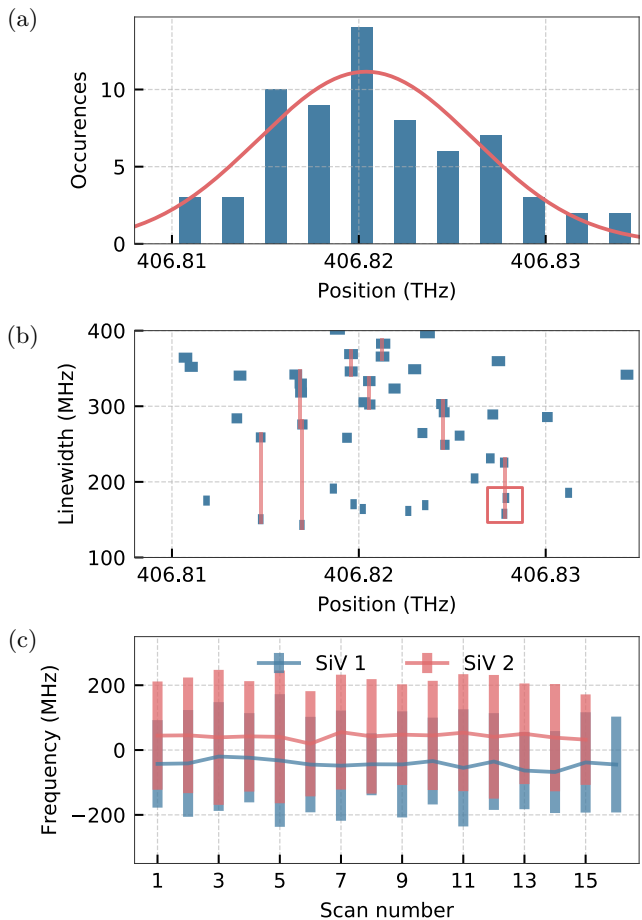
\begin{figure}[ht]
	\begin{tikzpicture}
	\draw (0,0) node[inner sep=0] {\includegraphics[scale=0.95]{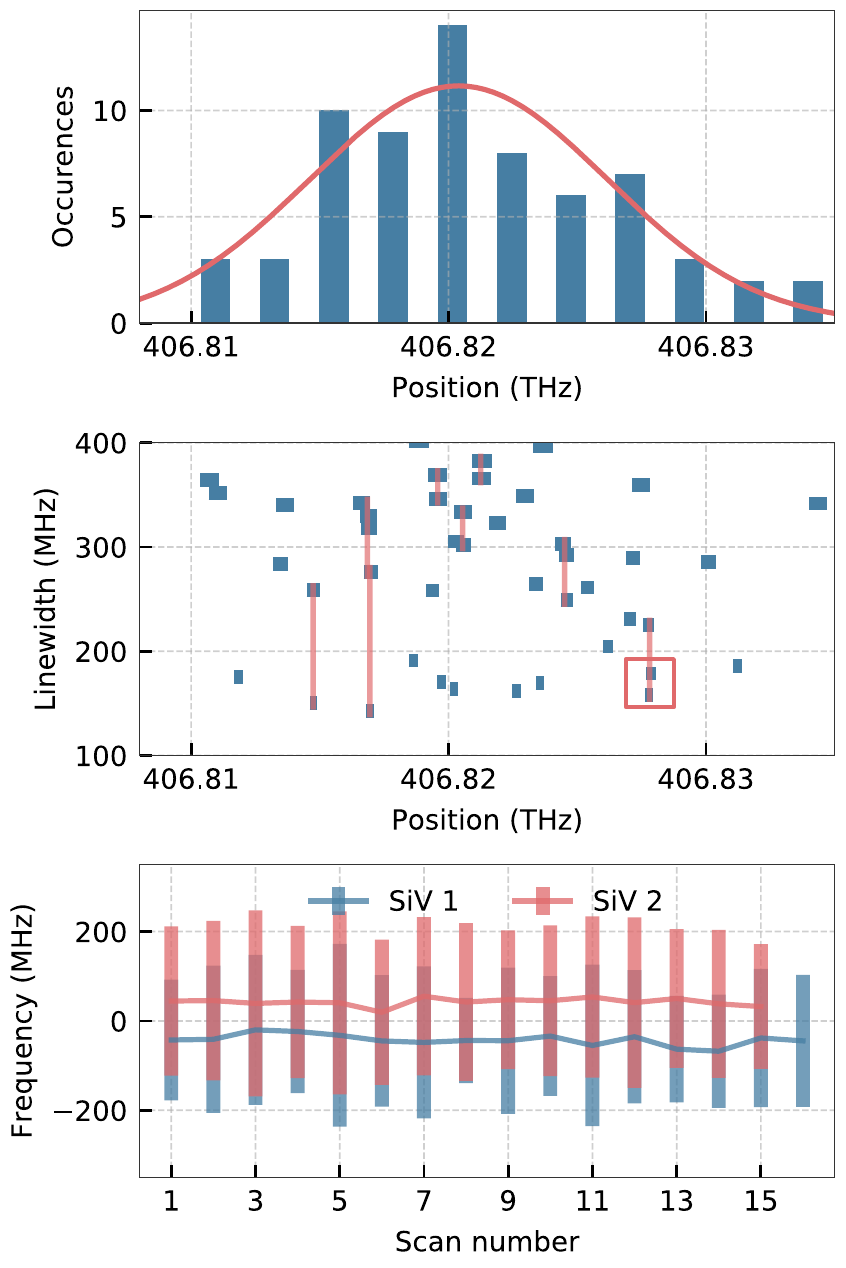}};
	\draw (-120pt, 170pt) node {(a)};
	\draw (-120pt, 50pt) node {(b)};
	\draw (-120pt, -65pt) node {(c)};
	\end{tikzpicture}
\caption{\label{fig:PLE_statistics}a) Histogram of the measured positions of transition C for different SiV$^-$s. The distribution has a width of 13.6 GHz and is centered at 406.82031 THz. b) Zoomed view into the distribution showing multiple possible transitions suitable for measuring two-photon interference. Emitters with significant overlap are marked with a red bar representing a linewidth of 94 MHz. The pair marked with the red square was used to show the two-photon interference. c) Evolution of the spectral position and linewidth for SiV 1 and SiV 2 over 15 scans, on a timescale of around 2 minutes. The center position is represented by the lineplot, the bars represent the FWHM of the individual scans.}
\end{figure}

The investigated SiV$^-$ show an inhomogeneous distribution with a full width at half maximum (FWHM) of around 14 GHz for the position of the C- transition, as shown in the histogram in FIG. \ref{fig:PLE_statistics} a). Multiple emitter show a significant overlap when comparing their spectral position, but some of the investigated lines show strong inhomogeneous broadening up to 1 GHz.
When setting an upper limit of 400 MHz for the inhomogeneous linewidth, we can still find multiple pairs and even groups of emitters which could be used for TPI. FIG. \ref{fig:PLE_statistics} b) depicts suitable candidates for TPI with linewidths below 400 MHz. The x-axis is the transition energy, while the y-axis shows the respective linewidth of an emitter. The width of the rectangular data points represents the linewidth of the measured emitter. Transitions with significant spectral overlap within the natural linewidth of 94 MHz \citep{Rogers2014} are connected with a red line. The two SiV$^-$ used for the TPI in the manuscript are marked by the red square. They were selected because of the combination of their spectral overlap, stability and narrow inhomogeneous linewidth. Under resonant excitation both SiV$^-$s were spectrally stable over a total of 15 scans, as shown in FIG. \ref{fig:PLE_statistics} c), where the spectral position is represented by the connecting lines and the respective linewidth by the colored bars. Also the overlapping region is depicted by the darker, shaded region.

\end{document}